# Screening of point charges in Si quantum dots

Alberto Franceschetti and M. Claudia Troparevsky

Oak Ridge National Laboratory, Oak Ridge, TN 37831[1]

**Abstract**

The screening of point charges in hydrogenated Si quantum dots ranging in diameter from 10 Å to 26 Å has been studied using first-principles density-functional methods. We find that the main contribution to the screening function originates from the electrostatic field set up by the polarization charges at the surface of the nanocrystals. This contribution is well described by a classical electrostatics model of dielectric screening.

---

The nature of dielectric screening in semiconductor quantum dots has been the subject of intense debate over the last several years. While there is a general agreement that the macroscopic dielectric constant is smaller in a quantum dot than in the corresponding bulk material, the physical origin of the size dependence of ε has been controversial [1-5]. The effects of the dot surface on the dielectric constant are often approximated using classical electrostatics models [6,7], even though the validity of such models becomes questionable when the quantum dot is only a few nanometers in size. According to classical electrostatics, an external charge in a finite dielectric body gives origin to polarization charges at the surface of the dielectric. These surface charges modify the electrostatic potential in the dielectric itself. As a result, the electrostatic field generated by the external charge is screened differently in a finite dielectric system than in an infinite dielectric with the same dielectric constant. For example, a point charge located at the center of a dielectric sphere of radius $R$ and dielectric constant $\varepsilon_{in}$ produces an electrostatic potential inside the sphere that is screened by a position-dependent effective dielectric constant:

$$\varepsilon_{eff}(r) = \frac{\varepsilon_{in}}{1 + \frac{r}{R}\left(\frac{\varepsilon_{in}}{\varepsilon_{out}} - 1\right)} \quad (r < R), \tag{1}$$

where $\varepsilon_{out}$ is the dielectric constant of the surrounding material. Classical electrostatics models based on Eq. (1) have been widely used in the literature to investigate the effects of surface polarization charges on the screening of electron-hole and electon-electron interactions in semiconductor quantum dots.

Recently, Ogut et al. [5] reported the first large-scale ab-initio calculations of the full dielectric function of hydrogenated Si clusters. They found a significant decrease in the dielectric constant of small clusters compared to bulk Si. Due to the complexity of such calculations, however, the authors considered only Si clusters containing up to 35 Si atoms. For such small clusters it is nearly impossible to separate the surface region from the interior region, so it is difficult to isolate the effects of the surface. Thus, the question of the validity of classical electrostatics models to describe screening in semiconductor quantum dots remains unanswered. In this work, we report first-principles density functional calculations of dielectric screening of point charges in Si quantum dots up to 26 Å in diameter (~500 Si atoms). Our calculations show that the screening of a point charge in a quantum dot is largely determined by the discontinuity of the dielectric constant at the surface of the quantum dot, as expected from classical electrostatics. Based on the results of our first-principles calculations, we propose a model screening function that can be readily applied to describe dielectric screening in semiconductor quantum dots of different materials, sizes and shapes.

The change in the electrostatic potential $\delta V(\mathbf{r})$ induced by an external, static charge distribution $\delta\rho_{ext}(\mathbf{r})$ can be written as:

$$\delta V(\mathbf{r}) = \int \frac{\delta\rho_{ext}(\mathbf{r}')}{\bar{\varepsilon}(\mathbf{r},\mathbf{r}')|\mathbf{r}-\mathbf{r}'|} d\mathbf{r}' \quad . \tag{2}$$

This equation *defines* the screening function $\bar{\varepsilon}(\mathbf{r},\mathbf{r}')$. If the charge distribution $\delta\rho_{ext}$

reduces to that of a point charge $q$ located at position $\boldsymbol{R}_0$, Eq. (2) becomes:

$$\delta V(\boldsymbol{r}) = \frac{q}{\bar{\varepsilon}(\boldsymbol{r}, \boldsymbol{R}_0) |\boldsymbol{r} - \boldsymbol{R}_0|} , \qquad (3)$$

which gives the test-charge test-charge screening function $\bar{\varepsilon}(\boldsymbol{r}, \boldsymbol{R}_0)$. The total electrostatic potential $\delta V(\boldsymbol{r})$ is the sum of the external potential $\delta V_{ext}(\boldsymbol{r}) = q/|\boldsymbol{r} - \boldsymbol{R}_0|$ and the induced electrostatic potential $\delta V_{ind}(\boldsymbol{r})$ due to the self-consistent rearrangement of the electronic charge density in response to the test charge $q$. Substituting $\delta V(\boldsymbol{r}) = \delta V_{ext}(\boldsymbol{r}) + \delta V_{ind}(\boldsymbol{r})$ into Eq. (3), we obtain the following closed expression for $\bar{\varepsilon}(\boldsymbol{r}, \boldsymbol{R}_0)$:

$$\bar{\varepsilon}(\boldsymbol{r}, \boldsymbol{R}_0) = \frac{1}{1 + |\boldsymbol{r} - \boldsymbol{R}_0| \delta V_{ind}(\boldsymbol{r})/q} . \qquad (4)$$

First-principles methods allow one to calculate $\delta V_{ind}(\boldsymbol{r})$ self consistently, and therefore to obtain $\bar{\varepsilon}(\boldsymbol{r}, \boldsymbol{R}_0)$ from Eq. (4). In principle, a separate self-consistent calculation is required for every position $\boldsymbol{R}_0$ of the test charge $q$. As a result, the calculation of the full screening function is computationally very time consuming [5]. We circumvent this problem by calculating $\bar{\varepsilon}(\boldsymbol{r}, \boldsymbol{R}_0)$ for selected positions of the test charge $q$.

Calculations are performed using density-functional theory in the local-density approximation (LDA). Ultra-soft pseudopotentials [8] are used to describe the interaction of the valence electrons with the ions. The quantum dot wave functions are expanded in a plane-wave basis set with an energy cutoff of 150 eV. We calculate the response of the system to external unit charges ($q=\pm e$) located at the positions of Si atoms. This is accomplished by replacing specific Si nuclei (atomic number Z) with a nucleus having one additional proton (Z+1) or a nucleus with one missing proton (Z-1). The total number of electrons is kept constant. In the presence of an external test charge, the cluster has a net charge of ±e, which is neutralized by introducing a uniform background charge of opposite sign. Electrostatic interactions between the cluster and its periodic images are corrected including up to quadrupolar terms. The induced potential $\delta V_{ind}(\boldsymbol{r})$ is then calculated by solving the Poisson equation:

$$\nabla^2 \delta V_{ind}(\boldsymbol{r}) = -4\pi \delta \rho_{ind}(\boldsymbol{r}) , \qquad (5)$$

where $\delta \rho_{ind}(\boldsymbol{r})$ is the change in the electronic charge density induced by the external charge $q$. The Poisson equation is solved by discretizing the Laplacian operator on a real space grid using a finite-differences approach. Boundary conditions are imposed by placing the passivated Si quantum dot in a large computational domain and evaluating the electrostatic potential at the boundaries of the domain from a multipole expansion [9]. The Poisson equation reduces then to a linear system that is solved iteratively using a conjugate-gradients minimization algorithm.

We consider here nearly spherical Si nanocrystals ranging in diameter from 10.4 Å to 26.0 Å. The inital atomic configurations of the Si nanocrystals are constructed by (i) cutting a spherical segment, centered on a Si atom, from a bulk Si crystal; (ii) removing surface atoms that have a coordination number lower than two; and (iii) passivating the dangling bonds of the remaining surface atoms by H atoms. The Si nanocrystal is then placed in a cubic supercell that is periodically repeated in space; the supercell is sufficiently large to avoid dot-dot interactions. In the next step, the Si and H atoms are randomly displaced (by less than 0.1 Å) to remove any remaining symmetries, and all the atomic positions are relaxed until quantum-mechanical forces acting on the ions are smaller than 0.01 eV/Å. We find that the Si nanocrystals have the $T_d$ symmetry in the equilibrium geometry. The nanocrystal sizes considered here are summarized in Table I.

| Nanocrystal | Effective radius (Å) | Supercell size (Å) |
|---|---|---|
| $Si_{29}H_{36}$ | 5.2 | 21.5 |
| $Si_{35}H_{36}$ | 5.6 | 21.5 |
| $Si_{87}H_{76}$ | 7.5 | 26.9 |
| $Si_{275}H_{112}$ | 11.0 | 32.3 |
| $Si_{275}H_{172}$ | 11.0 | 32.3 |
| $Si_{465}H_{228}$ | 13.0 | 37.7 |

Table I. Si nanocrystals considered in this work. The effective radius is defined in terms of the number of Si atoms in the nanocrystal.

The induced electronic charge $\Delta Q_{ind}(r) = \int_0^r \delta\rho_{ind}(r')\,dr'$, enclosed in a sphere of radius $r$ around a positive ($q=e$) test charge, is shown in Fig.1(a) for the $Si_{275}H_{172}$ and $Si_{275}H_{112}$ nanocrystals. The latter has a reconstructed surface obtained by forming Si-Si dimers on the <001>-oriented facets, and passivating the remaining Si dangling bonds with H atoms. In both cases the test charge is located at the center of the nanocrystal. We see from Fig. 1(a) that the induced charge is largely localized around the test charge, but has a significant component in the vicinity of the nanocrystal surface, as qualitatively predicted by classical electrostatics. Note that $\Delta Q_{ind}(r)$ integrates to zero, because the total number of electrons remains constant when the external charge is introduced. We also see from Fig.1(a) that the induced charge is nearly independent of surface passivation in the interior of the nanocrystal, while it depends on the type of passivation near the surface.

Figures 1(b) and 1(c) show the screening function $\bar{\varepsilon}(r)$ of $Si_{275}H_{172}$ (solid lines) and $Si_{275}H_{112}$ (dashed lines) nanocrystals, calculated for a positive test charge $q=+e$ [Fig. 1(b)], and for a negative test charge $q=-e$ [Fig. 1(c)]. Here the screening function, obtained from Eq. (4), is spherically averaged, and is plotted as a function of the distance $r$ from the test charge (located at the center of the nanocrystal). Note that $\bar{\varepsilon}(r)$ has a value of 1 at the origin ($r=0$), as dictated by Eq. (4) and by the fact that $\delta V_{ind}(r)$ is finite at the origin. $\bar{\varepsilon}(r)$ reaches a maximum around $r=1.1$ Å, and then decays slowly to 1 as the surface of the nanocrystal is approached. Remarkably, the $Si_{275}H_{172}$ and $Si_{275}H_{112}$ nanocrystals have nearly the same $\bar{\varepsilon}(r)$, notwithstanding the fact that their surface termination is very different. A comparison between Figs. 1(b) and 1(c) shows that, for r<2 Å, the screening function $\bar{\varepsilon}(r)$ depends on the sign of the test charge. In the linear-response regime, $\bar{\varepsilon}(r)$ does not depend (by definition) on the magnitude and sign of the test charge  The dependence of $\bar{\varepsilon}(r)$ on $q$ (see Fig. 1) indicates that the perturbation induced by a unit test charge exceeds the range of applicability of the linear-response approximation.

The spherically-averaged screening function $\bar{\varepsilon}(r/R)$ of Si nanocrystals of different sizes (see Table I) is shown in Fig. 2 as a function of the renormalized distance $r/R$ (where $R$ is the effective radius of the nanocrystals) from an external charge $q=-e$ located at the center of the nanocrystal. We see that $\bar{\varepsilon}(r/R)$ follows a universal curve (nearly independent of the nanocrystal radius $R$) for $r/R > 0.2$, while it depends on the nanocrystal size for $r/R < 0.2$. In particular, the maximum value of $\bar{\varepsilon}(r/R)$ increases as the radius of the nanocrystal increases. The screening function $\bar{\varepsilon}(r,r')$ of $Si_{35}H_{36}$ nanocrystals was calculated by Ogut et al. [5] using first-principles methods. $\bar{\varepsilon}(r,r')$ was derived form the inverse response function $\varepsilon^{-1}(r,r')$, calculated using ab-initio linear response theory. Our results are in good agreement with those of Ref. [5]. However, we find that the maximum value of $\bar{\varepsilon}(r/R)$ is higher in our calculations (3.9 vs. 2.6 for $Si_{35}H_{36}$). This is likely a consequence of the fact that the calculations of Ref. [5] were performed in the linear response approximation. Note also that the calculations of Ref. [5] included the exchange-correlation contribution to $\bar{\varepsilon}(r,r')$, which is omitted here since we are considering the test-charge test-charge screening function.

Figure 2 also shows the classical electrostatics screening function $\varepsilon_{eff}(r/R)$ given by Eq. (1). We used $\varepsilon_{in} = 11.9$ and $\varepsilon_{out} = 1$. Note that, since $\varepsilon_{eff}(r/R)$ does not depend explicitly on the nanocrystal radius $R$, it is a universal function of $r/R$. We see from Fig. 2 that $\varepsilon_{eff}(r/R)$ agrees remarkably well with $\bar{\varepsilon}(r/R)$ for $r/R > 0.2$, while it deviates from $\bar{\varepsilon}(r/R)$ at short range. The main source of this discrepancy is the fact that $\varepsilon_{eff}(r/R)$ does not approach the correct limit of 1 when $r/R \to 0$. This prompts us to propose the following model screening function, that has the correct asymptotic behavior at both short and long range:

$$\bar{\varepsilon}_{model}(r,r') = \frac{1}{\varepsilon_{in}} \varepsilon_{TF}(|r-r'|) \varepsilon_{eff}(r,r') \ . \tag{6}$$

Here $\varepsilon_{in}$ is the dielectric constant of bulk Si, and $\varepsilon_{TF}(|r-r'|)$ is the screening function of bulk Si in the Thomas-Fermi approximation [10]:

$$\varepsilon_{TF}(r) = \varepsilon_{in} \frac{q_{TF} R_{TF}}{\sinh[q_{TF}(R_{TF}-r)] + q_{TF} r}, \quad \text{for } r < R_{TF}$$

$$\varepsilon_{TF}(r) = \varepsilon_{in} \quad , \quad \text{for } r > R_{TF}. \tag{7}$$

The Thomas-Fermi wave vector $q_{TF}$ is given by $q_{TF} = 2\pi^{-1/2}(3\pi^2 n_0)^{1/3}$, where $n_0$ is the electron density, and the Thomas-Fermi radius $R_{TF}$ is the solution of the equation $sinh(q_{TF}$ $R_{TF})/(q_{TF} R_{TF}) = \varepsilon_{in}$. The last term on the rigt-hand side of Eq. (6), $\varepsilon_{eff}(r,r')$, is the classical effective screening function for a dielectric body of dielectric constant $\varepsilon_{in}$ embedded in a medium of dielectric constant $\varepsilon_{out}$. According to Eq. (3), $\varepsilon_{eff}(r,r')$ can be calculated as:

$$\varepsilon_{eff}(r,r') = \frac{q}{\phi(r)|r-r'|} \qquad (8)$$

where $\phi(r)$ is the solution of the generalized Poisson equation:

$$\nabla \cdot \varepsilon_M(r) \nabla \phi(r) = -4\pi q \delta(r-r') \cdot \qquad (9)$$

Here $\varepsilon_M(r)$ is the *macroscopic*, position-dependent dielectric constant of the system: $\varepsilon_M(r) = \varepsilon_{in}$ inside the quantum dot and $\varepsilon_M(r) = \varepsilon_{out}$ outside. Eq. (1) gives the expression of $\varepsilon_{eff}(r,r')$ in the particular case of a point charge located at the center ($r'=0$) of a dielectric sphere of radius $R$. Given the short-range character of $\varepsilon_{TF}(r)/\varepsilon_{in}$, which converges to 1 when $r \rightarrow R_{TF}$, the model screening function of Eq. (6) reduces to $\varepsilon_{eff}(r,r')$ for $|r-r'| > R_{TF}$.

In Fig. 3 we compare the LDA-derived screening function $\bar{\varepsilon}(r)$ of $Si_{87}H_{76}$ and $Si_{465}H_{228}$ nanocrystals with the model screening function $\bar{\varepsilon}_{model}(r)$ obtained by setting $r'=0$ in Eq. (6). Following Ref. [10], we use $q_{TF} = 1.1$ a.u. and $R_{TF} = 4.28$ a.u.. We find that the model screening function is in excellent agreement with the LDA screening function for $r > 0.2R$. For $r < 0.2R$, $\bar{\varepsilon}_{model}(r)$ understimates the effects of screening. We note, however, that the peak around $r = 1$ Å in the self-consistent calculation is expected to become less pronounced in the limit of an infinitesimal test charge (linear response regime), for which the model screening function of Eq. (6) has been derived. In addition, the deviation of $\bar{\varepsilon}_{model}(r)$ from the LDA screening function occurs in a relatively small volume (< 1% of the nanocrystal volume for R>10Å) around the test charge, so the effect on integrated quantities, such as electron binding energies, is expected to be small.

To test the accuracy of the model screening function, we have calculated the interaction energy $E_c$ of a positive test charge with an electron in the lowest conduction state (CBM):

$$E_c = \int |\psi_{CBM}(r)|^2 \frac{e^2}{\bar{\varepsilon}(r,R_0)|r-R_0|} dr , \qquad (10)$$

which corresponds, in first order approximation, to the binding energy of the electron with the positive charge. In Fig. 4 we compare the interaction energy $E_c$ calculated using the LDA screening fuction $\bar{\varepsilon}(r,R_0)$ of Eq. (4) and the model screening function $\bar{\varepsilon}_{model}(r,R_0)$ of Eq. (6). We see that $\bar{\varepsilon}_{model}(r,R_0)$ tends to slightly overestimate $E_c$ for small nanocrystal sizes, due to the smaller value of $\bar{\varepsilon}_{model}(r,R_0)$ in the vicinity of the test charge (Fig. 3). However, we see from Fig. 4 that the error in the interaction energy becomes negligible for R>10 Å , suggesting that Eq. (6) provides an excellent approximation to the screening function in this size regime.

In conclusion, we have reported first-principles, density-functional calculations of dielectric screening of point charges in hydrogenated Si nanocrystals ranging in size from 10 to 26 Å. We find that the main contribution to screening comes from the surface polarization charges induced by the discontinuity of the dielectric constant at the surface of the nanocrystal. We have proposed a model screening function [Eq. (6)] that can be applied to describe dielectric screening in semiconductor quantum dots as small as a few nanometers in diameter.

This work was supported by the U.S. Department of Energy, Office of Science, Basic Energy Sciences initiative LAB03-17. This research used resources of the National Energy Research Scientific Computing Center, which is supported by the Office of Science of the U.S. Department of Energy under Contract No. DE-AC03-76SF00098.

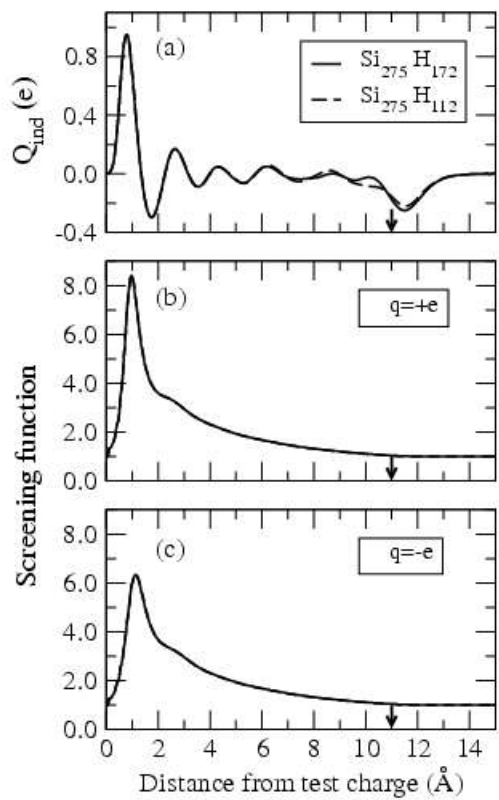

FIG. 1. Part (a) shows the electronic charge enclosed in a sphere of radius $r$ around a positive unit charge ($q=+e$) located at the center of $Si_{275}H_{172}$ (solid line) and $Si_{275}H_{112}$ (dashed line) nanocrystals. Parts (b) and (c) show the spherically averaged screening function $\bar{\varepsilon}(r)$ as a function of the distance $r$ from a positive (b) or negative (c) point charge. The vertical arrows denote the radius of the nanocrystal ($R = 11.0$ Å). Note that the screening functions of the $Si_{275}H_{172}$ and $Si_{275}H_{112}$ nanocrystals are almost indistinguishable.

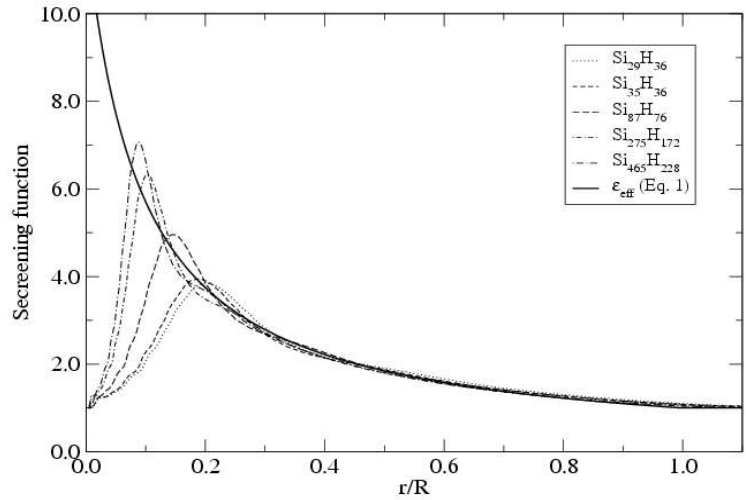

FIG. 2. Spherically averaged screening function $\bar{\varepsilon}(r/R)$ of several Si nanocrystals as a function of the rescaled distance ($r/R$) from the test charge ($q=-e$). Also shown (solid line) is the effective screening function calculated from Eq. (1) using $\varepsilon_{in}$ = 11.9 [Ref. 10] and $\varepsilon_{out}$ = 1.

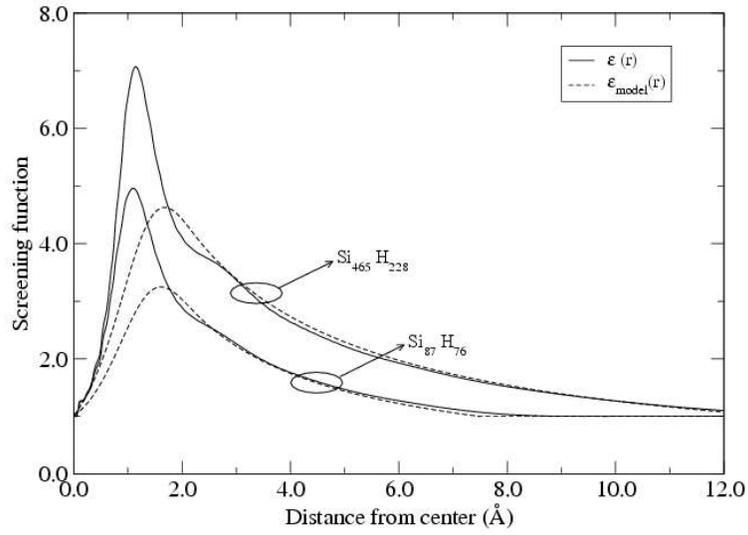

FIG. 3. Comparison between the LDA-derived screening function $\bar{\varepsilon}(r)$ (solid lines) and the model screening function $\bar{\varepsilon}_{\text{model}}(r)$ (dashed lines) for $Si_{87}H_{76}$ and $Si_{465}H_{228}$ nanocrystals.

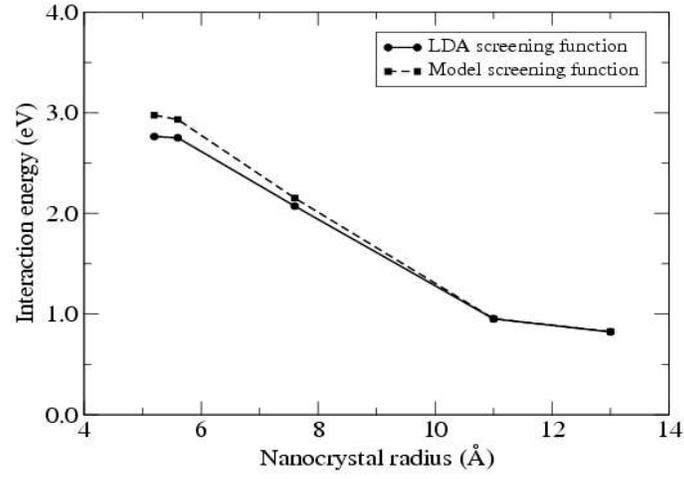

FIG. 4. Interaction energy between a positive unit charge at the center of the nanocrystal and an electron in the lowest unoccupied state, calculated using the LDA screening function $\bar{\varepsilon}(r, R_0)$ (solid line) and the model screening function $\bar{\varepsilon}_{model}(r, R_0)$ (dashed line).